\documentclass[aps,amssymb,preprint]{revtex4}                  
\usepackage{graphicx}
\usepackage{subfigure}
\newcommand{\e}{\epsilon}
\newcommand{\bb}{\beta}
\newcommand{\bc}{\beta_c}
\newcommand{\tr}{\tau}
\newcommand{\te}{\tau_{\rm exp}}

\newcommand{\cc}{c}
\newcommand{\ps}{p_{\rm swap}}

\newcommand{\de}{\beta \delta F}
\newcommand{\pp}{c}
\newcommand{\s}{\sigma}
\newcommand{\N}{K}
\newcommand{\sig}{\Delta}
\newcommand{\W}{{\cal W}}
\newcommand{\Ea}{{\cal E}_1}
\newcommand{\Eb}{{\cal E}_2}
\newcommand{\Ec}{{\cal E}_3}
\newcommand{\V}{{\cal V}}
\newcommand{\rr}{\rho}

\begin{document}

\title{Strengths and Weaknesses of Parallel Tempering
}

\author{J.~Machta}
\email{machta@physics.umass.edu}
\affiliation{
Physics Department,
University of Massachusetts,
Amherst, MA 010003 USA}

\begin{abstract}

Parallel tempering, also known as replica exchange Monte Carlo, is studied in the context of two simple free energy landscapes.  The first is a double well potential defined by two macrostates separated by a barrier.  The second is a `golf course' potential defined by microstates having two possible energies with exponentially more high energy states than low energy states. The equilibration time for replica exchange is analyzed for both systems.  For the double well system, parallel tempering with a number of replicas that scales as the square root of the barrier height yields exponential speedup of the equilibration time.  On the other hand, replica exchange yields only marginal speed-up for the golf course system.  For the double well system, the free energy difference between the two wells has a large effect on the equilibration time.  Nearly degenerate wells equilibrate much more slowly than strongly asymmetric wells.  It is proposed that this difference in equilibration time may lead to a bias in measuring overlaps in spin glasses. These examples illustrate the strengths and weaknesses of replica exchange and may serve as a guide for understanding and improving the method in various applications. 
\end{abstract}
\maketitle
\section{Introduction}
\label{sec:intro}

Replica exchange Monte Carlo (MC), also known as parallel tempering, was independently introduced by Swendsen and Wang~\cite{SwWa86} and Hukushima and Nemoto~\cite{HuNe96}  for studying spin glasses.   Replica exchange MC is an important tool in many areas of computational physics~\cite{EaDe05} where the free energy landscape has many metastable minima separated by barriers.  It is the standard method for simulating spin glasses and is used for protein folding~\cite{Ha97,ScHeVeWe05} and lattice gauge theory~\cite{BuFuKeMu07}.
 
In parallel tempering many replicas of the system are simulated in parallel using a standard MC technique for sampling the Gibbs distribution such as the Metropolis-Hastings algorithm.  The replicas are simulated at different temperatures. The fixed sequence of temperatures extends from some low temperature where the equilibration time  is very long to some high temperature where the equilibration time is short.  Replica exchange moves permit replicas at adjacent temperatures to swap temperatures in a way that satisfies detailed balance so that the entire set of replicas equilibrates at the prescribed set of temperatures.  The heuristic motivation for replica exchange is that replicas can diffuse from the lowest to the highest temperature and back to the lowest temperature.  During this `roundtrip' equilibration occurs at high temperature so that when the replica returns to the lowest temperature its state is independent of the original state.  A number of studies have focused on optimizing replica exchange MC by choosing the set of replica temperatures and other parameters to minimize the round-trip time~\cite{KaTrHuTr06,TrTrHa06, BiNuJa08}.  Replica exchange MC is also closely related both to simulated annealing and various generalized ensemble methods~\cite{Okamoto04}.

In the present paper we consider the efficiency of replica exchange MC in the context of two very simple free energy landscapes that, respectively, highlight the strengths and weaknesses of the method.  The first example is a free energy landscape with two minima separated by a barrier such as occurs in the $\phi^4$ free energy functional.  In the low temperature phase where the potential has a double well, local dynamics fully equilibrates in a time that is exponential in the barrier height though equilibration within a single well is typically much faster.   As we shall see, replica exchange can reduce the barrier crossing time from exponential to a polynomial in the barrier height.   

For the double well potential, we show that the equilibration time is longest when the free energy difference between the wells is a few $k_BT$ or less.  In this situation, the Gibbs measure gives significant weight to both macrostates and equilibration involves diffusive motion of replicas.  When the free energy difference between the wells is much larger than $k_BT$, equilibration depends on much faster ballistic motion of replicas.  In the context of spin glasses, a free energy landscapes with multiple free energy minima is expected.  In the discussion section, we argue that the longer equilibration time for nearly degenerate free energy minima may lead to an overestimate of the probability that the spin overlap in spin glasses is near zero. This reasoning suggests caution in interpreting simulations for the spin overlap. 

The second free energy landscape to be considered is the `golf course' potential.  Here almost all of the microstates are degenerate excited states and an exponentially small fraction of states are degenerate ground states. This model has a pseudo-first-order transition between a low temperature phase where the system is almost always in a ground state and a high temperature phase where the system is almost always in an excited state.  For this particularly nasty system, replica exchange is of little help equilibrating the low temperature phase.  The equilibration time is controlled by the time taken to find a ground state and this time is not reduced by bringing the system first to a higher temperature.

The outline of the paper is as follows.  In  Sec.\ \ref{sec:remc} we introduce replica exchange Monte Carlo Section.  In Sec.\ \ref{sec:dw} we analyze the behavior of replica exchange Monte Carlo for the double well free energy landscape both analytically and via numerical simulations.  In Sec.\ \ref{sec:golf} we introduce the golf course potential and analyze replica exchange for this potential.  The paper closes with a discussion in Sec.\ \ref{sec:disc}.

\section{Replica Exchange Monte Carlo}
\label{sec:remc}
Replica exchange Monte Carlo is designed to equilibrate a set of replicas of a system at inverse temperatures, 
\begin{equation}
\label{eq:repbb}
\bb_0 > \bb_1> \ldots, \bb_{R-1}
\end{equation}
Each of the $R$ replicas is equipped with dynamics that equilibrates it at its respective temperature.  In addition, replica exchange moves are permitted between replicas at neighboring temperatures. In a replica exchange move, the temperatures of the two replicas are swapped.  In order to satisfy detailed balance and insure that the entire set of replicas equilibrates, the probability for accepting a replica exchange move between replicas at temperatures $\bb$ and $\bb^\prime$ is given by
\begin{equation}
\label{eq:reprob}
\ps=\min\left[1, e^{(\bb-\bb^\prime)(E-E^\prime)}\right].
\end{equation}
where $E$ and $E^\prime$ are the respective energies of the replicas that were originally at $\bb$ and $\bb^\prime$.  If the replica exchange move is accepted, the replica whose local dynamics was set at inverse temperature $\bb$ is now set at $\bb^\prime$ and vice versa.   

\section{Double Well potential}
\label{sec:dw}
In this section we consider a simple free energy with two minima such as occurs, for example, in a $\phi^4$ theory.   The free energy associated with each minima or well is,
\begin{equation}
\label{eq:F}
\bb F_\s(\bb) = -\frac{1}{2}(\bb-\bc)^2 (\N+H\s)
\end{equation}
where $\s$ labels the well; the deep well is indicated by $\s=1$ and the shallow well  by $\s=0$. Although the free energy landscape itself is not explicitly prescribed, we assume that free energy at the saddle point between the wells is zero so that $F$ is the free energy barrier for transitions between wells.    From the free energy we can obtain the internal energy $U_\s(\bb)$ and energy fluctuations by differentiations with respect to $\bb$.    The internal energy, which is the expectation of the energy $E$ is
\begin{equation}
\label{eq:U}
U_\s(\bb)={\bf E}(E) = -(\bb-\bc) (\N+H\s)
\end{equation}
and the variance $\sig^2_\sigma$ of the energy is
\begin{equation}
\label{eq:sig}
\sig^2_\sigma={\bf Var}(E) = (\N+H\s)
\end{equation}
The free energy difference $\de(\bb)$ between the wells is controlled by $H$ ($H\geq 0$) and given by,
\begin{equation}
\label{eq:deltaE}
\de(\bb)= \bb F_0(\bb) - \bb F_1(\bb) = \frac{1}{2}(\bb-\bc)^2 H
\end{equation}

Given this free energy difference, the probability $\pp(\bb)$ of being in the deep well (i.\ e.\ the expectation of $\s$) at inverse temperature $\bb$ is
 \begin{equation}
\label{eq:pp}
\pp(\bb)={\bf E}(\sigma)=\frac{1}{1+e^{- \de(\bb)}}.
\end{equation}
To completely specify the statics of the model, we assume that the energy distribution in each well is a normal distribution with mean $U_\s(\bb)$ and variance $\sig^2_\sigma$. 

\subsection{Replica Exchange  Dynamics for the Double Well Model}
We suppose that each replica is equipped with single temperature dynamics that is much faster than the rate of replica exchange attempts.  Thus, for each replica exchange attempt, the energies of the two replicas are chosen independently from normal distributions for the given replica's temperature and well index.  The time scale for transitions between wells by single temperature dynamics for $\bb<\bc$ is order $e^{-\bb F}$, however, in the analysis and simulations that follow, we do not permit single temperature dynamics to effect transitions between the wells except at $\bc$.  This simplification leads to an underestimate of the equilibration time for replica exchange since the replicas near $\bc$ may also contribute to barrier crossings between the wells.   Since the entropy of each well is the same at $\bb=\bc$ and since there is no barrier between the wells here, we assume an initial condition for parallel tempering in which each well is equally likely to be populated for every $\bb<\bc$. 

It is straightforward to verify that the replica exchange dynamics described above satisfies detailed balance with respect to the normal distribution of the energies in the two wells and the probability $\pp(\bb)$ given in (\ref{eq:pp}) for being in the deep well.  The normal distribution of energies within each well is maintained by fiat while $\pp(\bb)$ is established via replica exchange.  Our goal is to understand the time scale for reaching this equilibrium well distribution.

Given the dynamical assumptions it is not difficult to calculate the average rate of replica exchange for the double well model.   Let $\W_{\sigma,\sigma^\prime}(\bb,\bb^\prime)$ be the average rate of replica exchange  if the two replicas are, respectively, at inverse temperatures $\bb$ and $\bb^\prime$ in wells $\sigma$ and $\sigma^\prime$.  Without loss of generality, assume $\bb\geq\bb^\prime$. The rate $\W_{\sigma,\sigma^\prime}(\bb,\bb^\prime)$ is obtained by averaging (\ref{eq:reprob}) over the energy distribution,
\begin{equation}
\label{ }
\W_{\sigma,\sigma^\prime}(\bb,\bb^\prime)={\bf E}( \min\left[1, e^{(\bb-\bb^\prime)(E-E^\prime)}\right]) .
\end{equation}
Here ${\bf E}(\cdot)$ is an average over the normal distributions of $E$ and $E^\prime$, the energies in the respective wells at the given temperatures. The explicit expression for the replica exchange rate is
\begin{eqnarray}
\label{eq:r}
\W_{\sigma,\sigma^\prime}(\bb,\bb^\prime)&=&\int\int \frac{dE \, dE^\prime}{2 \pi\sig_\sigma \sig_{\sigma^\prime}}e^{-(E-U_\s(\bb))^2/2\sig_\sigma^2-
(E^\prime-U_{\s^\prime}(\bb^\prime))^2/2\sig_{\sigma^\prime}^2} 
\\
\nonumber
&\times& \Bigg\{ \theta(E^\prime-E)e^{(\bb-\bb^\prime)(E-E^\prime)}+\theta(E-E^\prime)  \Bigg\}
\end{eqnarray}

\subsection{Degenerate wells: $H=0$}
\label{sec:h0}
First consider the simpler case of degenerate wells ($H=0$), it is not hard to show that the two terms in (\ref{eq:r}) are equal and, using (\ref{eq:U}) and (\ref{eq:sig}), the Gaussian integrals yield,
\begin{equation}
\label{eq:rate}
\W_{\sigma,\sigma^\prime}(\bb,\bb^\prime)= {\rm Erfc}\left( \frac{(\bb-\bb^\prime) \sqrt\N}{2}\right)
\end{equation}
where $ {\rm Erfc}(\cdot)$ is the complementary error function.  The replica exchange rates are independent of the well indices and the dynamics of replica exchange is diffusive.

Given the replica exchange rate, we can estimate the equilibration time as follows.  Suppose there are $R$ equally spaced replicas with inverse temperatures ranging from $\bb_0$ to $\bc$.   Equilibration requires that a replica in one well at the lowest temperature $\bb_0$ diffuses to $\bc$ where the well is randomized.  Up to constant factors, the equilibration time $\tr(R)$ for $R$ replicas scales like the mean first passage time for a random walk between the ends of a chain of $R$ sites with hopping rate $\W$, with a reflecting boundary at $\bb_0$ and an absorbing boundary at $\bc$.  The mean first passage time for this process is $(R-1)^2/\W$  \cite{redner01}.  Thus, from (\ref{eq:rate})
\begin{equation}
\label{eq:tauR}
\tr(R) \sim (R-1)^2/{\rm Erfc}\left(\frac{ (\bb_0-\bc) \sqrt\N}{2(R-1)}\right)
\end{equation} 
From the asymptotic behavior of the error function, ${\rm Erfc}(x) \approx \exp(-x^2)/(\sqrt\pi x)$ we see that the optimum number of replicas should scale as the square root of the well depth, $R_{\rm opt} \sim (\bb_0-\bc) \sqrt\N$.  The replica exchange rate is then order unity and the optimized equilibration time in this diffusive regime, $\tr^D$ is proportional to the well depth,
\begin{equation}
\label{eq:tauopt}
\tr^D \sim (R_{\rm opt}-1)^2 \sim K (\bb_0 -\bc)^2.
\end{equation}   
Since the optimum number of replicas is independent of prefactors in (\ref{eq:tauR}), we can obtain $R_{\rm opt}$ by numerically minimizing the RHS of (\ref{eq:tauR}) with respect to $R$, the result is 
\begin{equation}
\label{eq:ropt}
R_{\rm opt}=1+0.594 (\bb_0-\bc) \sqrt\N. 
\end{equation}

The above result represents the main strength of replica exchange Monte Carlo.  The time for barrier crossing between the wells has been reduced from an exponential in the barrier height to linear in the barrier height.  Note that a key feature of the double well model required for the success of parallel tempering is the continuity of the free energy with temperature.

\subsection{Asymmetric Wells: $H>0$}
\label{sec:h}
\begin{figure}
 \includegraphics{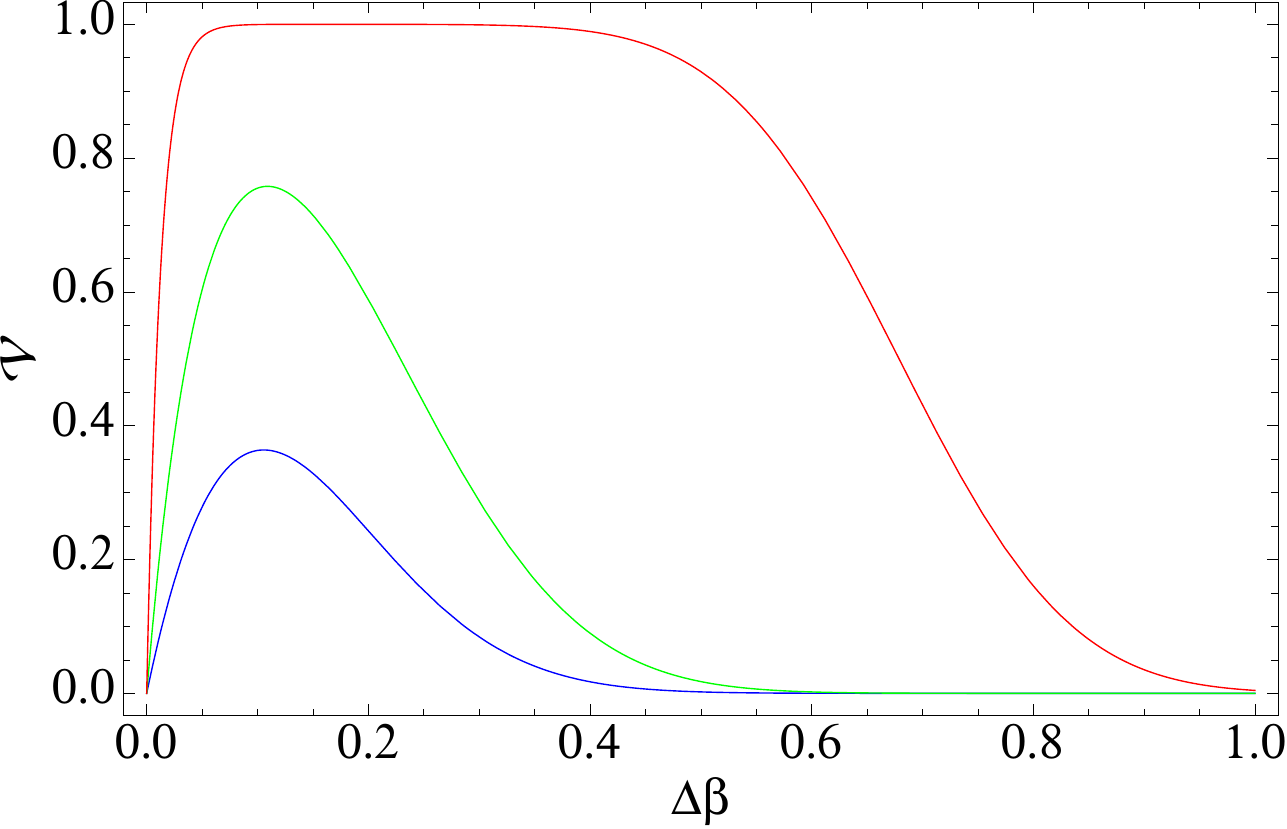}
\caption{The velocity $\V$ vs.\ the temperature between the replicas $\Delta \beta=\bb-\bb^\prime$ for three values of the well asymmetry parameter, $H=2$ (blue, bottom), 5 (green, middle) and 20 (red, top) for the choice $K=16$, $\bb=5$ and $\bc=1$.}
\label{fig:vel}       
\end{figure}

When $H>0$, the wells are asymmetric and the motion of replicas is biased diffusion.   Replicas in the deep well move toward lower temperatures relative to replicas in the shallow well.  The average replica exchange rates reflect that bias and, carrying out the integrals in (\ref{eq:r}), we obtain
\begin{equation}
\W_{0,1}(\bb,\bb^\prime)=\Ea+\Eb\Ec
\end{equation}
and
\begin{equation}
\W_{1,0}(\bb,\bb^\prime)=\Eb+\Ea/\Ec
\end{equation}
where
\begin{equation}
\Ea=\frac{1}{2}{\rm Erfc}\left( \frac{(\bb-\bb^\prime) \N - (\bb^\prime-\bc)H}{\sqrt{ 4\N+2H}}\right),
\end{equation}
\begin{equation}
\Eb=\frac{1}{2}{\rm Erfc}\left( \frac{(\bb-\bb^\prime) \N + (\bb-\bc)H}{\sqrt{ 4\N+2H}}\right),
\end{equation}
and
\begin{equation}
\Ec=\exp\left((\bb-\bb^\prime)(\frac{\bb+\bb^\prime}{2}-\bc) H\right).
\end{equation}
Since $\Ea>\Eb$ and $\Ec>1$, we have $\W_{0,1}(\bb,\bb^\prime)>\W_{1,0}(\bb,\bb^\prime)$ and the dynamics is biased toward deep well replicas moving to lower temperatures.  The velocity $\V(\bb,\bb^\prime)$ of deep well (shallow well) replicas  toward lower (higher) temperature is the difference of the rates,
\begin{equation}
\label{eq:V}
\V=\W_{0,1}-\W_{1,0}
\end{equation}
Figure \ref{fig:vel} shows $\V$ vs $\Delta \beta=\bb-\bb^\prime$ for several values of $H$ and the choice $K=16$, $\bb=5$ and $\bc=1$.  The three curves correspond to $H=2$ (bottom), $H=5$ (middle) and $H=20$ (top).  The qualitative features are that velocity increases as the bias, $H$ increases and that $\Delta \beta$ must neither be too large and nor too near zero to maximize the velocity.  As the bias increases, the velocity approaches unity for an increasing range of $\Delta \beta$.

As $H$ increases there is a complicated crossover from diffusion to biased diffusion to ballistic motion.  Unlike the symmetric case, for $H>0$ the arguments of the error functions depend on both the temperature difference between replicas and the absolute temperature so that evenly spaced  replicas cannot be expected to optimize the equilibration time.  Nonetheless, we can make some crude estimates for the ballistic regime.  The requirement for the ballistic regime is that  $\de=(\bb-\bc)^2H/2 \gg 1$ for most replica temperatures $\bb$ so that each replica is nearly always in the deep well.  Then $\Delta \bb$ can be chosen so that $\V \approx 1$.  In this regime the equilibration time scale is simply the time required to generate order $R$ states in the deep well at $\bc$ and then move them to the colder replicas.  This time scale is order $R$.  The argument of the error function must be small to achieve a velocity near unity, which is essentially the same condition as in the diffusive regime to that $R_{\rm opt} \sim (\bb_0-\bc) \sqrt\N$.  The behavior of the equilibration time in the ballistic regime $\tr^B$  is expected to be
\begin{equation}
\label{eq:taub}
\tr^B \sim R_{\rm opt} \sim (\bb_0-\bc) \sqrt\N .
\end{equation}
The main point here is that the time scale is much longer for the diffusive regime than the ballistic regime. In the diffusive regime, the equilibration time is proportional to the free energy barrier between the wells but in the ballistic regime it behaves as the square root of the free energy barrier.

We have so far considered the simple situation where the free energy difference between the wells changes monotonically with the temperature--one well is the deeper than the other for all $\bb>\bc$.  If instead, the free energy difference between wells changes sign in a temperature region where there is a large barrier between the wells then the motion of replicas will be biased in a way that causes trapping and very long equilibration times.  This is the situation that holds at a thermal first-order transition.   It would be interesting to calculate the equilibration times in a simple free energy landscape with a first-order transition.

\subsection{Numerical Simulations}
We carried out simulations of parallel tempering for the double well model.   In each step of the simulation, a pair of adjacent temperatures $\bb$ and $\bb^\prime$ is randomly chosen. The energies of the associated replicas are chosen from normal distributions with means $U_\s(\bb)$ and $U_{\s^\prime}(\bb^\prime)$ and standard deviations $\sig_\s$ and $\sig_{\s^\prime}$, respectively.  The replica exchange move is  accepted with probability given by  (\ref{eq:reprob}).   If one of the replicas is at $\bc$, the well indicator $\s$ for this replica is chosen randomly before attempting the replica exchange move.  Otherwise, transitions between the two wells are forbidden and $\s$ is conserved.  One MC sweep consists of $R-1$ replica exchange attempts and time is measured in sweeps.  The simulations are initialized so that each replica is randomly chosen to be in either well with equal probability.

We simulated several values of the parameters $H$, $\N$, and  $R$.   In all simulations we chose $\bb_0=5$ and $\bc=1$ so as to be deep in the low temperature regime.  We measured two quantities, the exponential autocorrelation time and the initial decay toward equilibrium.  The exponential autocorrelation time $\te$ is obtained from the  autocorrelation function, $\Gamma(t)$  of the fraction of replicas  in the deep well $\rr=(1/R)\sum_{i=0}^{R-1} \s_i$, 
\begin{equation}
\label{eq:defrho}
\Gamma(t)= \frac{\langle \rr(t) \rr \rangle -\langle \rr \rangle^2}
{\langle \rr^2 \rangle -\langle \rr \rangle^2}.
\end{equation}
Here $\langle \cdot \rangle$ indicates an equilibrium average and the measurement of $\rr(t)$ is displaced by $t$ sweeps from $\rr$.  The equilibrium average is obtained from long time averages. The initialization time before data collection is typically several thousand sweeps and the run time is 10 to 50 million sweeps.  Except for an initial period of faster decay, $\Gamma(t)$ is well described by a single exponential and the exponential autocorrelation time $\te$ is obtained by fitting   to a single exponential function, $\Gamma(t) = a e^{-t/\te}$ over an appropriate range of $t$.    We considered $\N=8$, 16, 32 and 64 and $H=0$, 0.5, 1, 2, 3, 5 and 10.  We explored a range of numbers of replicas $R$ to find $R_{\rm opt}$.  We found that $R_{\rm opt}$ is correctly predicted by (\ref{eq:ropt}) for $H=0$ and that for $H>0$, $R_{\rm opt}$ is slightly larger than for the symmetric $H=0$ case but within one or two of the predictions of (\ref{eq:ropt}). An exact measurement of $R_{\rm opt}$ for the asymmetric case proved difficult because $\tau(R)$ varies very little with $R$ near $R_{\rm opt}$. 

\begin{figure}
\includegraphics{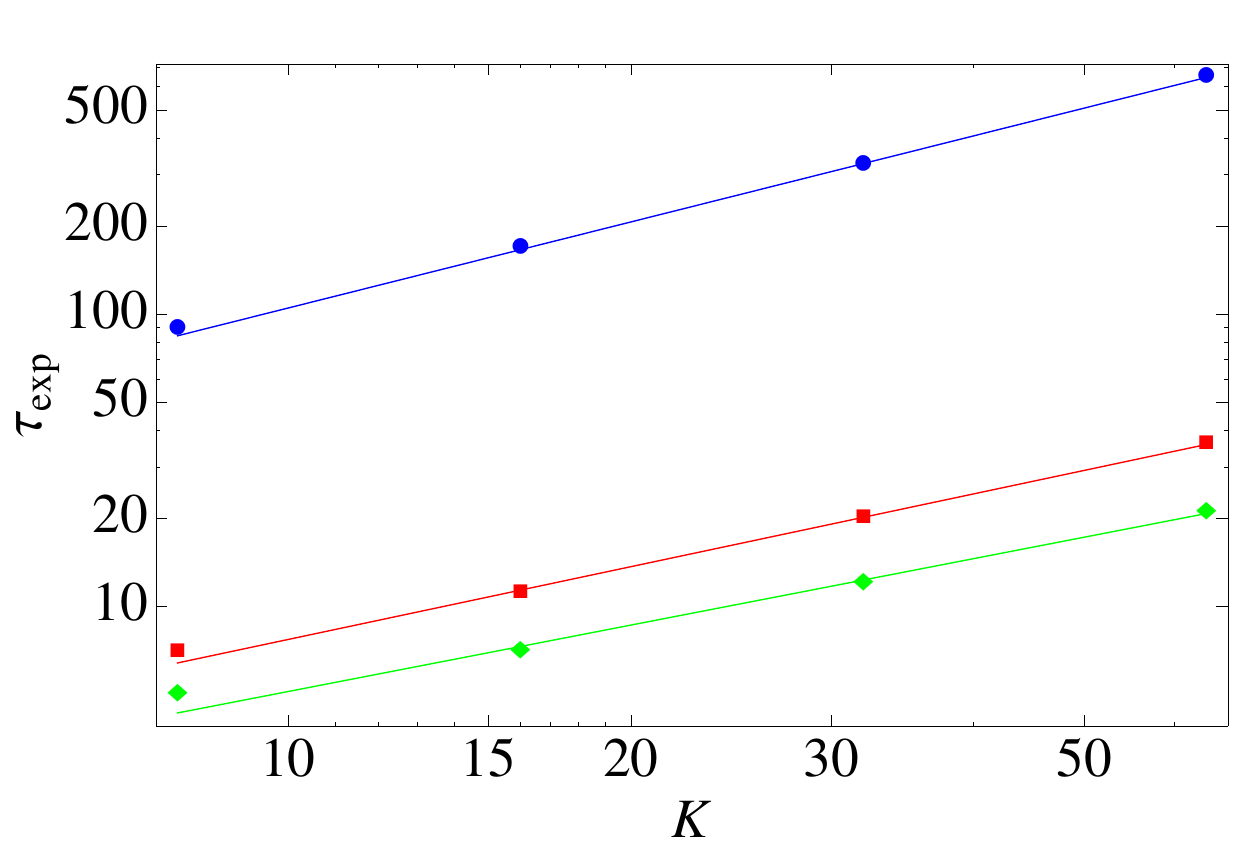}
\caption{The exponential autocorrelation time $\te$ for the fraction of sites in the deep well vs.\ the well depth parameter $\N$ for $H=10$ (green diamonds), $H=5$ (red squares) and $H=0$ (blue circles).  The lines are best power law fits, $\te \sim \N^x$ with $x=0.76$, 0.83 and 0.99 for $H=10$, 5 and 0, respectively.}
\label{fig:texpvk}       
\end{figure}

\begin{figure}
 \includegraphics{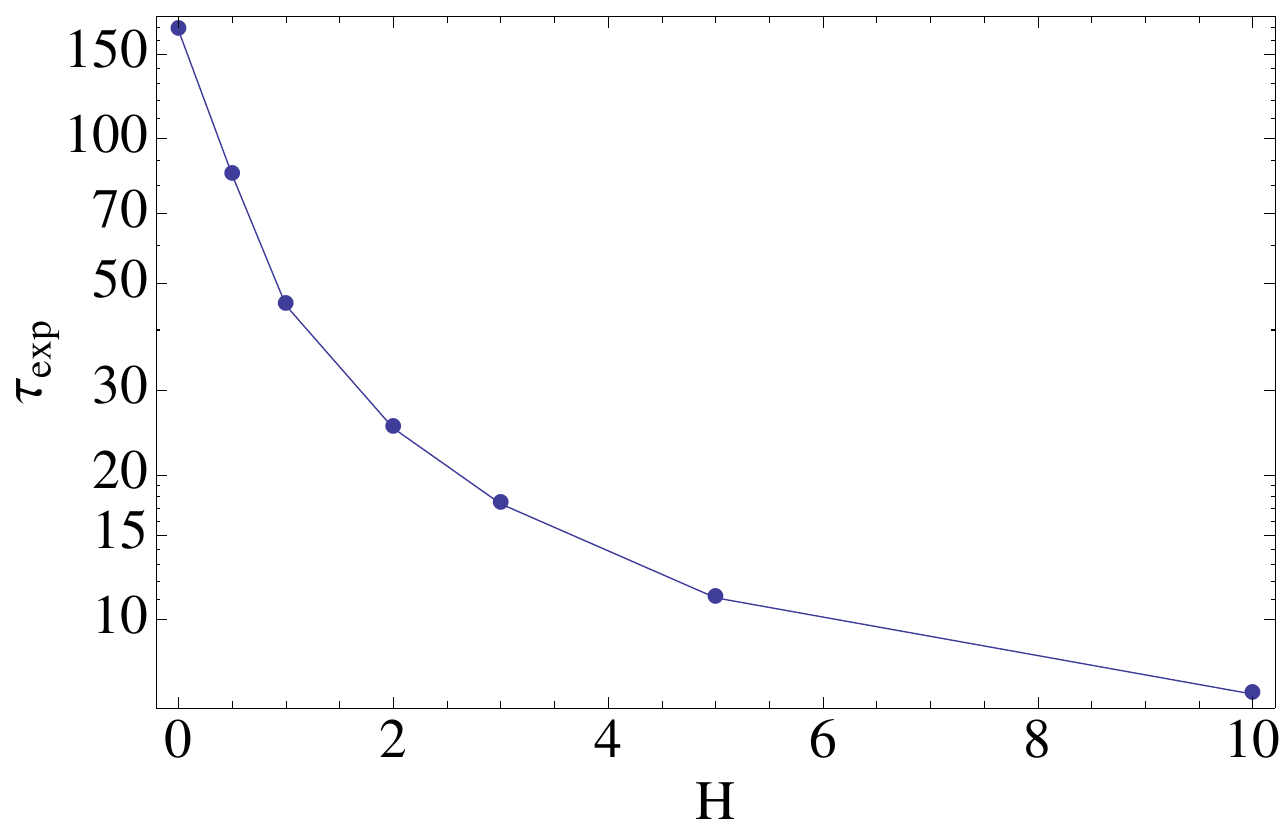}
\caption{The exponential autocorrelation time $\te$ vs.\ the asymmetry between the wells $H$ for $K=16$.}
\label{fig:tauvH}      
 \end{figure}

Figure \ref{fig:texpvk} shows $\te$ vs.\ $\N$ for $H=0$, 5 and 10 and for $\N=8$, 16, 32 and 64. Statistical errors are smaller than the data points.  The lines are best power law fits  of the form $a\N^x$.  The fitted powers are $x=0.76$, 0.83 and 0.99 for $H=10$, 5 and 0, respectively. The fact that $\te$ increases essentially linearly in $\N$ for $H=0$ agrees with (\ref{eq:tauopt}) of Sec.\ \ref{sec:h0} for symmetric wells.  However, although $x<1$ for the two asymmetric cases, we do not observe $x=0.5$, as predicted in (\ref{eq:taub}) of Sec.\ \ref{sec:h} for highly asymmetric wells, even for $H=10$.  We believe this is a crossover effect but it may also indicate more subtleties in the asymmetric case than have been taken into account in the simple theoretical arguments based on ballistic motion of replicas.  
Figure \ref{fig:tauvH} shows $\te$ vs $H$ for fixed $K=16$ revealing the rapid decline in equilibration time as the asymmetry increases.
  
\begin{figure}
\begin{center}
 \subfigure[ $H=0.1$]{\includegraphics{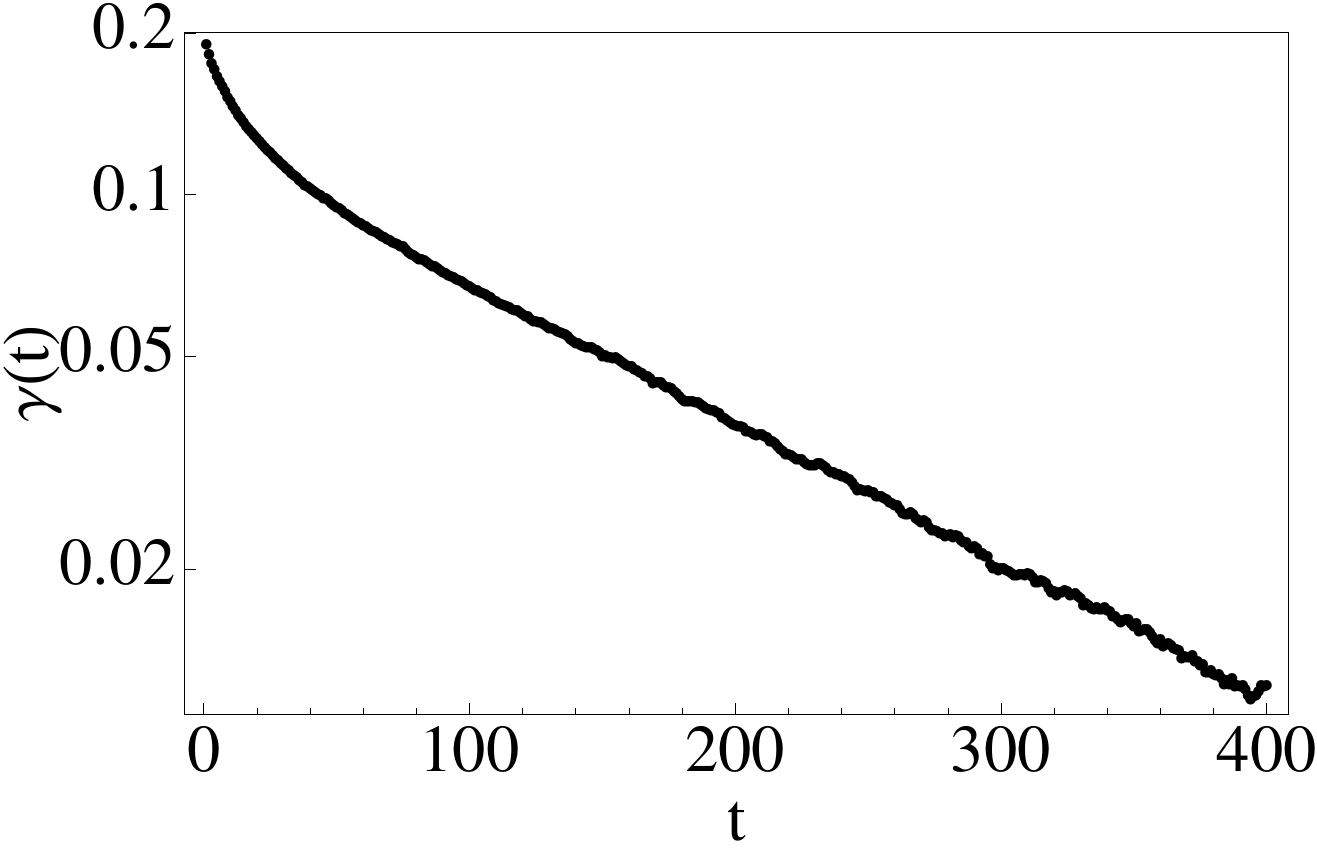}}
\subfigure[ $H=5.0$]{ \includegraphics{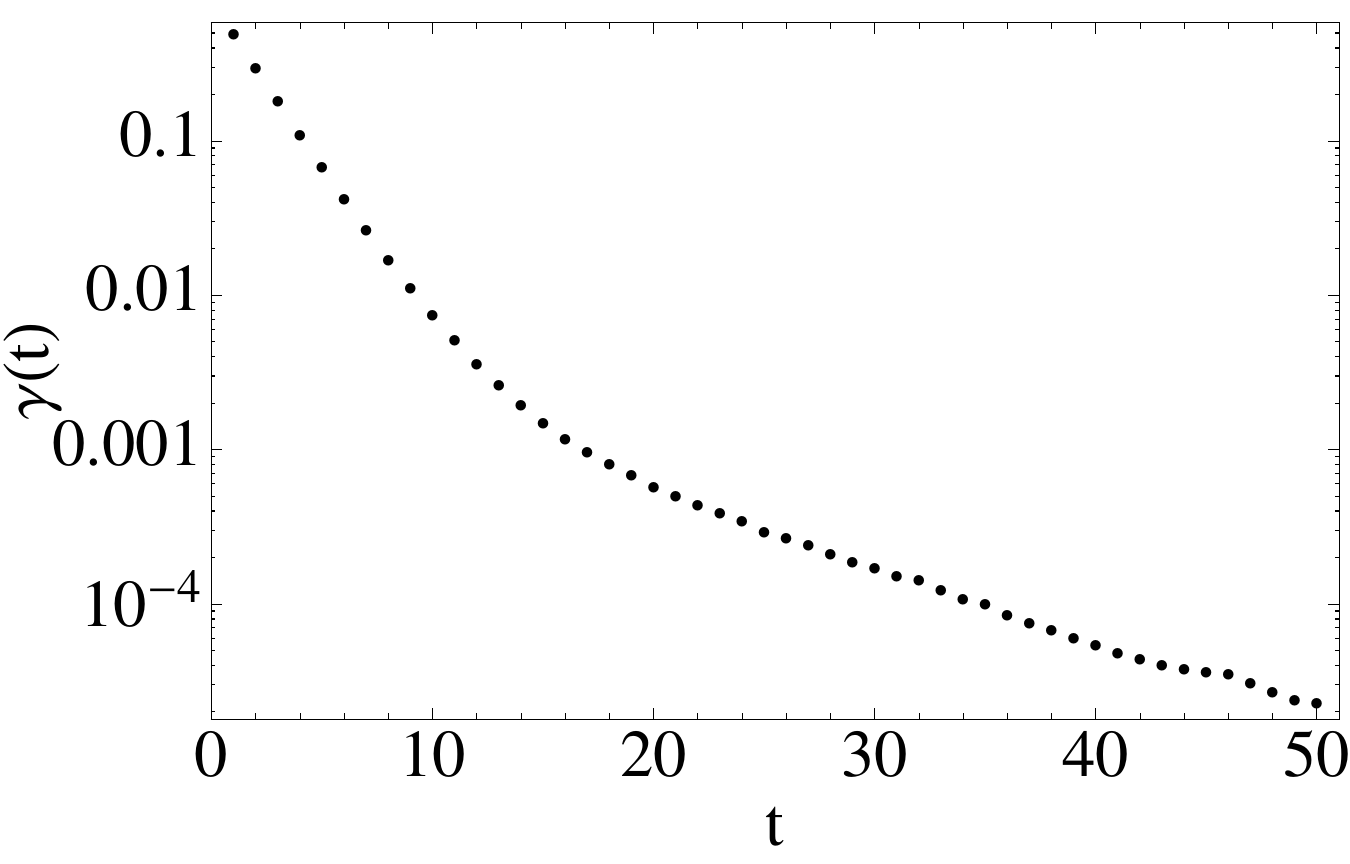}}
\end{center}
\caption{The initial approach to equilibrium of the probability of the coldest replica being in the deep well $\gamma(t)$ vs time $t$ after the initial conditions for (a) small asymmetry  $H=0.1$  and (b) large asymmetry $H=5.0$. In both cases $\N=16$, $\bb_0=5$ and  $R=12$.}
\label{fig:decayH01}       \end{figure}

The initial relaxation to equilibrium of the lowest temperature replica is often more relevant for applications of parallel tempering than the exponential autocorrelation time.  To study the initial relaxation to equilibrium, we investigated $\gamma(t)$, the probability that the lowest temperature replica is in the deep well after $t$ sweeps,  
\begin{equation}
\gamma(t)=\langle \s_0(t) \rangle - \pp(\bb_0).
\end{equation}
Note that $\langle \s_{0}(0) \rangle=1/2$ for the standard initial condition and $\lim_{t\rightarrow\infty}\langle \s_0(t) \rangle=\pp(\bb_0)$ so that $\gamma(t)$ approaches zero for large $t$.  The error in sampling $\s_0$ after $t$ sweeps is determined by $\gamma(t)$.  Figures  \ref{fig:decayH01} 
shows loglinear plots of  $\gamma(t)$ as a function of $t$ for (a) small asymmetry $H=0.1$  and (b) large asymmetry $H=5$  for $K=16$ and $R=12$.  For small asymmetry, $\gamma(t)$ decays nearly exponentially after an initial faster decay with a time scale that is very close to $\te$.
For large asymmetry, two time scales are clearly apparent.  The short time scale is approximately 2 and the long time scale is about 9, whereas $\te=11$ for these parameters.  Presumably, the  time scale $\te$ would finally be apparent in $\gamma(t)$ but perhaps not until it has decayed to extremely small values.  

The initial short time scale for $\gamma(t)$ for the strongly asymmetric $H=5$ case can be understood qualitatively as the average time for the lowest temperature replica that is also in the deep well to move to the lowest temperature $\bb_0$. If this nearest replica is at temperature $\bb_k$ the expected time for it to move to $\bb_0$ is approximately $k/\V$ where $\V$ is the velocity defined in (\ref{eq:V}), which is nearly unity in the strongly asymmetric case for sufficiently closely spaced replicas.    Thus we expect that initially, $\gamma(t)$ will decay on a time scale order unity.  Since this initial decay persists to quite small values of $\gamma(t)$ it may be the relevant time scale for practical equilibration in the strongly asymmetric case.

The key finding of the numerical study is that equilibration times are much shorter for asymmetric free energy minima than for nearly degenerate minima.

\section{Golf course landscape}
\label{sec:golf}
In this section we consider the efficacy of replica exchange for the `golf course' landscape.  Like a flat golf course, this landscape has a small number of degenerate ground states--the `hole'--and an exponentially larger number of degenerate excited states--the `green.'  The golf-course landscape has $e^N$ microstates where $N$ is the `system size.'  A fraction $e^{-\beta_c \e N}$ of these microstates have energy 0 while the remaining states have energy $\e N$ with $\e>0$.  The golf course system is a quenched disordered system; different realizations of disorder correspond to different sets of ground states. The microstates of the system are labeled by integers and we suppose that there is an oracle for each realization of the system that tells whether a given integer label corresponds to a ground state.  On the other hand, the set of ground states is itself inaccessible except by exhaustive search.  Natural realizations of golf course landscapes are studied in \cite{Baum86}.

The parameter $\bc$ is also the inverse temperature of a pseudo-first-order transition.  For $\bb > \bc$ and large $N$ the system is almost surely in a ground state but for $\bb < \bc$ it is almost surely in an excited state.  More specifically, let $\cc(\bb)$ be the probability of being in a ground state at temperature $\bb$.  It is straightforward to see that $\cc(\bb)$ is given by
\begin{equation}
\label{ }
\cc(\bb)= \frac{1}{1+e^{-(\bb-\bc)\e N}}.
\end{equation}

The dynamics of the system is a global version of the Metropolis-Hastings algorithm.  Each step consists of proposing a random microstate, consulting the oracle to determine its energy and then accepting the proposal with probability, $\min\left[1,e^{-\bb \Delta E}\right]$ where $\Delta E$ is the difference in energy between the proposed and initial microstate. This dynamics converges to equilibrium, however, the equilibration time is exponential in $N$.  In units of MC steps, the excitation rate from a ground state to an excited state is controlled by the energy barrier  $e^{-\bb \e N}$.  The de-excitation rate from an excited state to a ground state is controlled by the entropy barrier is $e^{-\bc \e N}$.  The equilibration rate is the sum of these two rates, $e^{-\bc \e N}+e^{-\bb \e N}$.  For  $\bb_0 > \bc$ the equilibration rate is dominated by the de-excitation term and the equilibration time $\tau_0$ is approximately,
\begin{equation}
\label{eq:tau0}
\tau_0 \approx e^{-\bc \e N}
\end{equation}
On the other hand, for $\bb_1<\bc$, the equilibration rate is dominated by the excitation rate and the equilibration time $\tau_1$ is 
\begin{equation}
\label{eq:tau1}
\tau_1 \approx e^{-\bb_1 \e N}
\end{equation}

Suppose we wish to equilibrate a system at inverse temperature $\bb_0>\bc$ using replica exchange Monte Carlo with replicas given by (\ref{eq:repbb}) with $\bb_{R-1} \leq \bc$.  Initially each replica is almost certainly in an excited state.  Since all replicas have the same energy, replica exchange moves are always accepted and the round trip time is independent of $N$ though proportional to $R^2$.  Of course, this short initial round trip time is misleading and has nothing to do with the equilibration time.  Equilibration requires finding ground states and this happens on the exponential time scale $\tau_0$.  If a ground state is discovered by a high temperature replica, that ground state will quickly and nearly irreversibly be passed to lower temperature by replica exchange.  If there are $\ell$ replica temperatures in the low temperature `phase,' that is, if $\bb_{\ell-1} < \bc$ and $\bb_\ell > \bc$ then a ground state must be discovered $\ell$ times to populate each cold replica.  Since there are $R$ systems looking for the ground state, we obtain a modest acceleration of order $R /\ell$.  For example, if $\ell = 1$, there is a factor of $R$ speed-up due to replica exchange.  This speed-up is not dependent on faster equilibration at high temperatures but relies on simple parallelism; all replicas are put to work independently looking for rare ground states but only $\ell$ ground states need to be found.    In conclusion, for the golf course landscape, replica exchange achieves a modest speed-up in the equilibration time due to brute force parallelism.  Before equilibration has been achieved, the round-trip time is short and unrelated to the equilibration time.  The golf course landscape is the most extreme case of the problem of the equilibrium macrostate having an exponentially small and hidden basin of attraction.

\section{Discussion}
\label{sec:disc}
The broad conclusion of this work is that replica exchange Monte Carlo is efficient for systems with free energy landscapes with multiple  minima so long as the landscape varies continuously and monotonically with temperature and the relevant minima have large basins of attraction.  In this situation, replica exchange is able to quickly sample the minima with the correct weighting.  In this context, a basin of attraction of a macrostate  is imprecisely defined as the subset of microstates that has a high likelihood of reaching the macrostate via a quench from high temperature.  The speed of the quench must be much slower than the rate of equilibration within the macrostate and much faster than the transition rate between macrostates.  Given the assumption of large basins of attraction and exponential barriers between macrostates, parallel tempering with polynomially many replicas reaches equilibrium in a time that is polynomial in the barrier height and thus achieves exponential speed-up.  On the other hand, replica exchange yields little improvement for systems where the relevant macrostates states have small basins of attraction.  Here the problem is simply finding equilibrium states rather than moving between them.  For real world applications, both kinds of problems may be present--barriers between multiple states and small basins of attraction.  

Let's now consider the case of Ising spin glasses in three dimensions.  It is is known that finding ground states is NP-hard~\cite{Bara}.  This fact suggest, but does not prove, that the basins of attraction of the low temperature equilibrium states are exponentially small.  On the other hand, for temperatures not too far below the critical temperature it may be that the basins of attraction are still relatively large and replica exchange can produce large reductions in equilibration times.  

As a working hypothesis, let us adopt the droplet picture~\cite{FiHu86, BrMo87a} for the low temperature phase of the three-dimensional Ising spin glass.  Within the droplet picture  something like the double well model should describe the lowest lying states in the low temperature phase.  The two wells correspond to the two orientations of the droplet and fluctuations around these orientations\footnote{Each droplet state is actually a degenerate pair related by a global spin flip.}.  Each realization of disorder has different values of the barrier height and free energy difference.  The statistics of these parameters are assumed to have power law behavior in the linear systems size $L$. Specifically, $\overline{\N}\sim L^\psi$ and $\overline{H^2}\sim L^{2\theta}$ where the overbar refers to a disorder average and $\theta$ is believed to be near 0.2 for the three-dimensional Ising spin glass.  

Parallel tempering has proved to be a successful tool for studying 3D spin glasses and it is reasonable to assume that for small systems the equilibrium states have sufficiently large basins of attraction that they can be ``found" in a reasonable time by replica exchange.  Even so, there is a potential source of bias in parallel tempering as it is typically used.  In typical applications  the length of the run is fixed independent of the realization of disorder.  These parameters are chosen to insure that some disorder averages are near their equilibrium values.  For example, the test described in \cite{KaPaYo01} insures that the disorder averaged energy is near its equilibrium value.  However, this requirement may not guarantee that all relevant observables are well equilibrated.   

In the droplet model, the fraction of realizations with nearly degenerate lowest lying states scales as $L^{-\theta}$.   Thus most disorder realizations have a large free energy difference between the equilibrium state and the excited (droplet) state.  In the context of the double well model, these realizations have $(\bb-\bc)^2H/2 \gg 1$ and, as we have seen, they will be rapidly equilibrated by parallel tempering.   On the other hand, the small fraction (order $L^{-\theta}$) of realizations with `active droplets,' that is, two nearly degenerate minima,  will have much longer than typical equilibration times.  Both because these realizations are rare and because the two droplet orientations have similar energies, the poor equilibration of these active droplet realizations will not introduce much error in the measurement of the disorder averaged energy. The same cannot be said for the disorder averaged spin overlap distribution near zero overlap, $P(q\approx0)$.  It is precisely the difficult to equilibrate, active droplet realizations that contribute to this quantity since these are the systems that have a significant likelihood in equilibrium of having either droplet orientation.  If these realizations are not equilibrated it will lead to an overestimate of $P(q\approx0)$. In particular, the equilibrium contribution of a given realization to $P(q\approx0)$ depends on the relative weight of the two droplet states.  Given the simplifying assumption that the two droplet states have zero overlap,  a realization with free energy difference $\de(\bb)$ between the droplet states will contribute $2\pp(\bb)(1-\pp(\bb))=2 e^{-\de}/(1+e^{-\de})^2$ to $P(q\approx0)$.  Exactly degenerate disorder realizations contribute $1/2$ to $P(q\approx0)$ but as $\de$ becomes larger than unity, the equilibrium contribution to $P(q\approx0)$ diminishes rapidly. However, for times less than the equilibration time the two droplet orientations will be close to equally populated assuming both have nearly equal basins of attraction as expected in the droplet model.  The result of this bias is that, $P(q\approx0)$ approaches its equilibrium value {\em from above} on a time scale associated with the equilibration of realizations with active droplets.  This time scale is expected to be considerably longer than the time scale for the equilibration of the disorder averaged energy.

The nature of the low temperature phase of finite-dimensional spin glasses is the subject of a long standing controversy.  In the droplet scenario most realizations of disorder have a large gap between a unique equilibrium state and a macroscopically different low lying excited state.  The replica symmetry breaking picture~\cite{Parisi79, MePaVi,MaPaRu98,CoGiGiPaVe07,Parisi08} proposes multiple nearly degenerate low lying equilibrium states with an ultrametric overlap structure.  A scenario that has features of both pictures and is supported by numerical studies is the `trivial non-trivial' (TNT) picture~\cite{KrMa00,PaYo00}.   The possible state structures in the thermodynamic limit are constrained by mathematical theorems~\cite{NeSt02} but the various scenarios are difficult to distinguish in simulations of small systems.  The disorder averaged spin overlap near zero,  $P(q\approx0)$ has been studied numerically to distinguish these scenarios~\cite{PaYo00,KrMa00, MaZu99}.  In the RSB and also the TNT an pictures, this quantity approaches a constant while in the droplet picture it decreases as $L^{-\theta}$.  In simulations $P(q\approx0)$ decreases for small $L$ but then reaches a plateau at a small but nonzero value \cite{KaYo02}.   The above considerations suggest caution in interpreting numerical results for $P(q\approx0)$.  It would be useful to carefully study the correlation between the equilibration time of disorder realizations and their contribution to $P(q\approx0)$ to insure that this quantity has been correctly measured in simulations.
 
\begin{acknowledgments}
I thank Matthias Troyer, Michael Moore, Burcu Yucesoy and Peter Young for useful discussions.
\end{acknowledgments}


\begin{thebibliography}{26}
\expandafter\ifx\csname natexlab\endcsname\relax\def\natexlab#1{#1}\fi
\expandafter\ifx\csname bibnamefont\endcsname\relax
  \def\bibnamefont#1{#1}\fi
\expandafter\ifx\csname bibfnamefont\endcsname\relax
  \def\bibfnamefont#1{#1}\fi
\expandafter\ifx\csname citenamefont\endcsname\relax
  \def\citenamefont#1{#1}\fi
\expandafter\ifx\csname url\endcsname\relax
  \def\url#1{\texttt{#1}}\fi
\expandafter\ifx\csname urlprefix\endcsname\relax\def\urlprefix{URL }\fi
\providecommand{\bibinfo}[2]{#2}
\providecommand{\eprint}[2][]{\url{#2}}

\bibitem[{\citenamefont{Swendsen and Wang}(1986)}]{SwWa86}
\bibinfo{author}{\bibfnamefont{R.~H.} \bibnamefont{Swendsen}} \bibnamefont{and}
  \bibinfo{author}{\bibfnamefont{J.-S.} \bibnamefont{Wang}},
  \bibinfo{journal}{Phys. Rev. Lett.} \textbf{\bibinfo{volume}{57}},
  \bibinfo{pages}{2607} (\bibinfo{year}{1986}).

\bibitem[{\citenamefont{Hukushima and Nemoto}(1996)}]{HuNe96}
\bibinfo{author}{\bibfnamefont{K.}~\bibnamefont{Hukushima}} \bibnamefont{and}
  \bibinfo{author}{\bibfnamefont{K.}~\bibnamefont{Nemoto}},
  \bibinfo{journal}{J. Phys. Soc. Jpn.} \textbf{\bibinfo{volume}{65}},
  \bibinfo{pages}{1604} (\bibinfo{year}{1996}).

\bibitem[{\citenamefont{Earl and Deem}(2005)}]{EaDe05}
\bibinfo{author}{\bibfnamefont{D.~J.} \bibnamefont{Earl}} \bibnamefont{and}
  \bibinfo{author}{\bibfnamefont{M.~W.} \bibnamefont{Deem}},
  \bibinfo{journal}{Phys. Chem. Chem. Phys.} pp. \bibinfo{pages}{3910--3916}
  (\bibinfo{year}{2005}).

\bibitem[{\citenamefont{Hansmann}(1997)}]{Ha97}
\bibinfo{author}{\bibfnamefont{U.~H.~E.} \bibnamefont{Hansmann}},
  \bibinfo{journal}{Chem. Phys. Lett.} \textbf{\bibinfo{volume}{281}},
  \bibinfo{pages}{140} (\bibinfo{year}{1997}).

\bibitem[{\citenamefont{Schug et~al.}(2005)\citenamefont{Schug, Herges, Verma,
  and Wenzel}}]{ScHeVeWe05}
\bibinfo{author}{\bibfnamefont{A.}~\bibnamefont{Schug}},
  \bibinfo{author}{\bibfnamefont{T.}~\bibnamefont{Herges}},
  \bibinfo{author}{\bibfnamefont{A.}~\bibnamefont{Verma}}, \bibnamefont{and}
  \bibinfo{author}{\bibfnamefont{W.}~\bibnamefont{Wenzel}},
  \bibinfo{journal}{Journal of Physics: Condensed Matter}
  \textbf{\bibinfo{volume}{17}}, \bibinfo{pages}{S1641} (\bibinfo{year}{2005}).

\bibitem[{\citenamefont{Burgio et~al.}(2007)\citenamefont{Burgio, Fuhrmann,
  Kerler, and Muller-Preussker}}]{BuFuKeMu07}
\bibinfo{author}{\bibfnamefont{G.}~\bibnamefont{Burgio}},
  \bibinfo{author}{\bibfnamefont{M.}~\bibnamefont{Fuhrmann}},
  \bibinfo{author}{\bibfnamefont{W.}~\bibnamefont{Kerler}}, \bibnamefont{and}
  \bibinfo{author}{\bibfnamefont{M.}~\bibnamefont{Muller-Preussker}},
  \bibinfo{journal}{Physical Review D} \textbf{\bibinfo{volume}{75}},
  \bibinfo{eid}{014504} (\bibinfo{year}{2007}).

\bibitem[{\citenamefont{Katzgraber et~al.}(2006)\citenamefont{Katzgraber,
  Trebst, Huse, and Troyer}}]{KaTrHuTr06}
\bibinfo{author}{\bibfnamefont{H.~G.} \bibnamefont{Katzgraber}},
  \bibinfo{author}{\bibfnamefont{S.}~\bibnamefont{Trebst}},
  \bibinfo{author}{\bibfnamefont{D.~A.} \bibnamefont{Huse}}, \bibnamefont{and}
  \bibinfo{author}{\bibfnamefont{M.}~\bibnamefont{Troyer}},
  \bibinfo{journal}{J. Stat. Mech.} \bibinfo{eid}{03018}
  (\bibinfo{year}{2006}).

\bibitem[{\citenamefont{Trebst et~al.}(2006)\citenamefont{Trebst, Troyer, and
  Hansmann}}]{TrTrHa06}
\bibinfo{author}{\bibfnamefont{S.}~\bibnamefont{Trebst}},
  \bibinfo{author}{\bibfnamefont{M.}~\bibnamefont{Troyer}}, \bibnamefont{and}
  \bibinfo{author}{\bibfnamefont{U.~H.~E.} \bibnamefont{Hansmann}},
  \bibinfo{journal}{J. Chem. Phys.} \textbf{\bibinfo{volume}{124}},
  \bibinfo{pages}{174903} (\bibinfo{year}{2006}).

\bibitem[{\citenamefont{Bittner et~al.}(2008)\citenamefont{Bittner,
  Nu{\ss}baumer, and Janke}}]{BiNuJa08}
\bibinfo{author}{\bibfnamefont{E.}~\bibnamefont{Bittner}},
  \bibinfo{author}{\bibfnamefont{A.}~\bibnamefont{Nu{\ss}baumer}},
  \bibnamefont{and} \bibinfo{author}{\bibfnamefont{W.}~\bibnamefont{Janke}},
  \bibinfo{journal}{Phys. Rev. Lett.} \textbf{\bibinfo{volume}{101}},
  \bibinfo{pages}{130603} (\bibinfo{year}{2008}).

\bibitem[{\citenamefont{Okamoto}(2004)}]{Okamoto04}
\bibinfo{author}{\bibfnamefont{Y.}~\bibnamefont{Okamoto}},
  \bibinfo{journal}{Journal of Molecular Graphics and Modelling}
  \textbf{\bibinfo{volume}{22}}, \bibinfo{pages}{425 } (\bibinfo{year}{2004}).

\bibitem[{\citenamefont{{Fisher} and {Huse}}(1986)}]{FiHu86}
\bibinfo{author}{\bibfnamefont{D.~S.} \bibnamefont{{Fisher}}} \bibnamefont{and}
  \bibinfo{author}{\bibfnamefont{D.~A.} \bibnamefont{{Huse}}},
  \bibinfo{journal}{Phys. Rev. Lett.} \textbf{\bibinfo{volume}{56}},
  \bibinfo{pages}{1601} (\bibinfo{year}{1986}).

\bibitem[{\citenamefont{Bray and Moore}(1987)}]{BrMo87a}
\bibinfo{author}{\bibfnamefont{A.~J.} \bibnamefont{Bray}} \bibnamefont{and}
  \bibinfo{author}{\bibfnamefont{M.~A.} \bibnamefont{Moore}}, in
  \emph{\bibinfo{booktitle}{Heidelberg Colloquium on Glassy Dynamics}}, edited
  by \bibinfo{editor}{\bibfnamefont{L.}~\bibnamefont{Van~Hemmen}}
  \bibnamefont{and}
  \bibinfo{editor}{\bibfnamefont{I.}~\bibnamefont{Morgenstern}}
  (\bibinfo{publisher}{Springer Verlag}, \bibinfo{year}{1987}), vol.
  \bibinfo{volume}{275} of \emph{\bibinfo{series}{Lecture Notes in Physics}},
  p.~\bibinfo{pages}{57}.

\bibitem[{\citenamefont{Redner}(2001)}]{redner01}
\bibinfo{author}{\bibfnamefont{S.}~\bibnamefont{Redner}},
  \emph{\bibinfo{title}{A Guide to First-Passage Processes}}
  (\bibinfo{publisher}{Cambridge University Press},
  \bibinfo{address}{Cambridge}, \bibinfo{year}{2001}).

\bibitem[{\citenamefont{Baum}(1986)}]{Baum86}
\bibinfo{author}{\bibfnamefont{E.~B.} \bibnamefont{Baum}},
  \bibinfo{journal}{Phys. Rev. Lett.} \textbf{\bibinfo{volume}{57}},
  \bibinfo{pages}{2764} (\bibinfo{year}{1986}).

\bibitem[{\citenamefont{Barahona}(1982)}]{Bara}
\bibinfo{author}{\bibfnamefont{F.}~\bibnamefont{Barahona}},
  \bibinfo{journal}{J. Phys. A: Math. Gen.} \textbf{\bibinfo{volume}{15}},
  \bibinfo{pages}{3241} (\bibinfo{year}{1982}).

\bibitem[{\citenamefont{{Katzgraber} et~al.}(2001)\citenamefont{{Katzgraber},
  {Palassini}, and {Young}}}]{KaPaYo01}
\bibinfo{author}{\bibfnamefont{H.~G.} \bibnamefont{{Katzgraber}}},
  \bibinfo{author}{\bibfnamefont{M.}~\bibnamefont{{Palassini}}},
  \bibnamefont{and} \bibinfo{author}{\bibfnamefont{A.~P.}
  \bibnamefont{{Young}}}, \bibinfo{journal}{Phys. Rev. B}
  \textbf{\bibinfo{volume}{63}}, \bibinfo{pages}{184422}
  (\bibinfo{year}{2001}).

\bibitem[{\citenamefont{Parisi}(1979)}]{Parisi79}
\bibinfo{author}{\bibfnamefont{G.}~\bibnamefont{Parisi}},
  \bibinfo{journal}{Phys. Rev. Lett.} \textbf{\bibinfo{volume}{43}},
  \bibinfo{pages}{1754} (\bibinfo{year}{1979}).

\bibitem[{\citenamefont{Mezard et~al.}(1987)\citenamefont{Mezard, Parisi, and
  Virasoro}}]{MePaVi}
\bibinfo{author}{\bibfnamefont{M.}~\bibnamefont{Mezard}},
  \bibinfo{author}{\bibfnamefont{G.}~\bibnamefont{Parisi}}, \bibnamefont{and}
  \bibinfo{author}{\bibfnamefont{M.}~\bibnamefont{Virasoro}},
  \emph{\bibinfo{title}{Spin Glass Theory and Beyond, Spin Glass Theory and
  Beyond}}, vol.~\bibinfo{volume}{9} of \emph{\bibinfo{series}{World Scientific
  Lecture Notes in Physics}} (\bibinfo{publisher}{World Scientific Publishing
  Company}, \bibinfo{year}{1987}).

\bibitem[{\citenamefont{{Marinari} et~al.}(1998)\citenamefont{{Marinari},
  {Parisi}, and {Ruiz-Lorenzo}}}]{MaPaRu98}
\bibinfo{author}{\bibfnamefont{E.}~\bibnamefont{{Marinari}}},
  \bibinfo{author}{\bibfnamefont{G.}~\bibnamefont{{Parisi}}}, \bibnamefont{and}
  \bibinfo{author}{\bibfnamefont{J.~J.} \bibnamefont{{Ruiz-Lorenzo}}},
  \bibinfo{journal}{Phys. Rev. B} \textbf{\bibinfo{volume}{58}},
  \bibinfo{pages}{14852} (\bibinfo{year}{1998}).

\bibitem[{\citenamefont{Contucci et~al.}(2007)\citenamefont{Contucci, Giardina,
  Giberti, Parisi, and Vernia}}]{CoGiGiPaVe07}
\bibinfo{author}{\bibfnamefont{P.}~\bibnamefont{Contucci}},
  \bibinfo{author}{\bibfnamefont{C.}~\bibnamefont{Giardina}},
  \bibinfo{author}{\bibfnamefont{C.}~\bibnamefont{Giberti}},
  \bibinfo{author}{\bibfnamefont{G.}~\bibnamefont{Parisi}}, \bibnamefont{and}
  \bibinfo{author}{\bibfnamefont{C.}~\bibnamefont{Vernia}},
  \bibinfo{journal}{Phys. Rev. Lett.} \textbf{\bibinfo{volume}{99}},
  \bibinfo{pages}{057206} (\bibinfo{year}{2007}).

\bibitem[{\citenamefont{Parisi}(2008)}]{Parisi08}
\bibinfo{author}{\bibfnamefont{G.}~\bibnamefont{Parisi}}, \bibinfo{journal}{J.
  Phys. A: Math. Gen.} \textbf{\bibinfo{volume}{41}}, \bibinfo{pages}{324002}
  (\bibinfo{year}{2008}).

\bibitem[{\citenamefont{Krzakala and Martin}(2000)}]{KrMa00}
\bibinfo{author}{\bibfnamefont{F.}~\bibnamefont{Krzakala}} \bibnamefont{and}
  \bibinfo{author}{\bibfnamefont{O.~C.} \bibnamefont{Martin}},
  \bibinfo{journal}{Phys. Rev. Lett.} \textbf{\bibinfo{volume}{85}},
  \bibinfo{pages}{3013} (\bibinfo{year}{2000}).

\bibitem[{\citenamefont{Palassini and Young}(2000)}]{PaYo00}
\bibinfo{author}{\bibfnamefont{M.}~\bibnamefont{Palassini}} \bibnamefont{and}
  \bibinfo{author}{\bibfnamefont{A.~P.} \bibnamefont{Young}},
  \bibinfo{journal}{Phys. Rev. Lett.} \textbf{\bibinfo{volume}{85}},
  \bibinfo{pages}{3017} (\bibinfo{year}{2000}).

\bibitem[{\citenamefont{Newman and Stein}(2001)}]{NeSt02}
\bibinfo{author}{\bibfnamefont{C.~M.} \bibnamefont{Newman}} \bibnamefont{and}
  \bibinfo{author}{\bibfnamefont{D.~L.} \bibnamefont{Stein}},
  \bibinfo{journal}{J. Stat. Phys.} \textbf{\bibinfo{volume}{106}},
  \bibinfo{pages}{213} (\bibinfo{year}{2001}).

\bibitem[{\citenamefont{Marinari and Zuliani}(1999)}]{MaZu99}
\bibinfo{author}{\bibfnamefont{E.}~\bibnamefont{Marinari}} \bibnamefont{and}
  \bibinfo{author}{\bibfnamefont{F.}~\bibnamefont{Zuliani}},
  \bibinfo{journal}{J. Phys. A: Math. Gen.} \textbf{\bibinfo{volume}{32}},
  \bibinfo{pages}{7447} (\bibinfo{year}{1999}).

\bibitem[{\citenamefont{Katzgraber and Young}(2002)}]{KaYo02}
\bibinfo{author}{\bibfnamefont{H.~G.} \bibnamefont{Katzgraber}}
  \bibnamefont{and} \bibinfo{author}{\bibfnamefont{A.~P.} \bibnamefont{Young}},
  \bibinfo{journal}{Phys. Rev. B} \textbf{\bibinfo{volume}{65}},
  \bibinfo{pages}{214402} (\bibinfo{year}{2002}).

\end{thebibliography}
\end{document}